\documentclass{elsarticle}

\usepackage{lineno,hyperref}
\usepackage{framed,multirow}
\usepackage{booktabs}
\usepackage{amssymb}
\usepackage{latexsym}
\usepackage{algorithm, algorithmic}
\usepackage{url}
\usepackage{xcolor}
\usepackage{amsmath}
\usepackage[shortlabels]{enumitem}
\modulolinenumbers[5]

\begin{document}

\begin{frontmatter}

\title{Building Lightweight Semantic Segmentation Models for Aerial Images Using Dual Relation Distillation}





\author[mymainaddress]{Minglong Li}
\ead{liminglong18@mails.ucas.ac.cn}

\author[mymainaddress]{Lianlei Shan}
\ead{shanlianlei18@mails.ucas.ac.cn}

\author[mymainaddress]{Weiqiang Wang\corref{mycorrespondingauthor}}
\cortext[mycorrespondingauthor]{Corresponding author}
\ead{wqwang@ucas.ac.cn}

\author[mysecondaryaddress]{Ke Lv}
\ead{luk@ucas.ac.cn}

\author[mythirdaddress]{Bin Luo}
\ead{luobin@ahu.edu.cn}

\author[mythirdaddress]{Si-Bao Chen}
\ead{sbchen@ahu.edu.cn}

\address[mymainaddress]{School of Computer Science and Technology,  University of Chinese Academy of Sciences, Beijing, China}

\address[mysecondaryaddress]{School of Engineering Science, University of Chinese Academy of Sciences, Beijing, China}

\address[mythirdaddress]{MOE Key Lab of Signal Processing and Intelligent Computing, School of Computer Science and Technology, 
Anhui University, Hefei, China}

\begin{abstract}
{Recently, there have been significant improvements in the accuracy of CNN models for semantic segmentation. However, these models are often heavy and suffer from low inference speed, which limits their practical application. To address this issue, knowledge distillation has emerged as a promising approach to achieve a good trade-off between segmentation accuracy and efficiency. In this paper, we propose a novel dual relation distillation (DRD) technique that transfers both spatial and channel relations in feature maps from a cumbersome model (teacher) to a compact model (student). Specifically, we compute spatial and channel relation maps separately for the teacher and student models, and then align corresponding relation maps by minimizing their distance. Since the teacher model usually learns more information and collects richer spatial and channel correlations than the student model, transferring these correlations from the teacher to the student can help the student mimic the teacher better in terms of feature distribution, thus improving the segmentation accuracy of the student model. We conduct comprehensive experiments on three segmentation datasets, including two widely adopted benchmarks in the remote sensing field (Vaihingen and Potsdam datasets) and one popular benchmark in general scene (Cityscapes dataset). The experimental results demonstrate that our novel distillation framework can significantly boost the performance of the student network without incurring extra computational overhead.}
\end{abstract}



\end{frontmatter}


\section{Introduction}
\label{sec:intro}
{Semantic segmentation is a crucial image understanding task that aims to perform per-pixel categorizations for a given image. In recent years, effective methods based on Fully Convolutional Networks (FCNs) \cite{Long2014Fully} have significantly enhanced the accuracy of segmentation. However, the high computational complexity of these models poses a challenge for their efficient deployment on resource-constrained devices, such as edge or IoT devices.

One promising solution to address this challenge is to design novel neural architectures or operators that are more efficient on edge systems without compromising accuracy. For instance, Sandler et al.  \cite{Sandler_2018_CVPR} add linear bottleneck and inverted residual block to improve both accuracy and performance. Zhou et al. \cite{10.1007/978-3-030-58580-8_40} combines neural architecture search and NetAdapt to design a more accurate and efficient network architecture. 
Ma et al.\cite{Ma_2018_ECCV} empirically observes four principles for designing efficient CNNs and proposes channel split to improve accuracy and performance.
Recently, \cite{ORSIC2021107611} presents pyramidal fusion, a principled approach for dense recognition based on resolution pyramids, which outperforms all previous semantic segmentation approaches aiming at real-time operation.

\par In addition to designing computationally efficient architectures, model compression is another popular approach to achieve a trade-off between accuracy and speed. Generally, existing compression methods can be roughly divided into three categories: quantization, pruning, and knowledge distillation. Quantization-based approaches represent the parameters of filter kernels and weighting matrices using fewer bits. Pruning-based methods aim to trim the network by removing redundant connections between neurons of adjacent layers. Knowledge distillation is a way of transferring knowledge from a cumbersome model to a compact model to improve the performance of compact networks.

\par The notion of knowledge distillation was first proposed in \cite{2006Model}. 
In the computer vision field, knowledge distillation was initially applied to image classification tasks by using the class probabilities produced from the cumbersome model as soft targets for training the compact model  \cite{10.5555/2969033.2969123} or further transferring the intermediate feature maps \cite{Zagoruyko2017AT}. Some recent works have extended knowledge distillation to semantic segmentation tasks. Since the goal of semantic segmentation is to perform per-pixel categorizations for a given image, a straightforward idea to introduce knowledge distillation into semantic segmentation tasks is to align pixel-wise outputs individually. This procedure forces the student model to mimic the teacher model in terms of output probability maps.

However, unlike classification tasks, semantic segmentation involves structured output, where correlations between a certain pixel and other pixels can help accurately classify that pixel. Therefore, correlations between different pixels are crucial for accurate segmentation, especially when the distance between two pixels exceeds the receptive field. The teacher and student models usually capture different long-range contextual information due to their differences in receptive fields. Thus, transferring this information from the teacher to the student helps the student model achieve better segmentation results. 
Liu et al.\cite{Liu_2019_CVPR} present a method to distill structured knowledge from large networks to small networks, Wang et al. \cite{wang2020ifvd} focus on transferring intra-class feature variation from teacher to student, and Li et al.\cite{9328142} propose a novel dynamic-hierarchical attention distillation network with multimodal synergetic instance selection for land cover classification using missing data modalities. Among these methods,\cite{Liu_2019_CVPR} and \cite{wang2020ifvd}  have made use of spatial relations in feature maps and achieved good results.

Moreover, several works have paid attention to channel relation. \cite{s20164616} models channel and spatial correlation in a single matrix for both teacher and student, then aligns the two matrices using a channel and spatial correlation loss function. 
Shu et al.\cite{shu2020cwd} apply the softmax normalization to transfer the activation values in each channel into a distribution. They then directly minimize the discrepancy of the channel-wise distribution between the teacher and student network. For a given feature map, contextual relationships not only exist in the spatial dimension but also in the channel dimension. The work 
\cite{shu2020cwd} ignores the relation between different channels and implement channel distillation separately for each channel. In contrast, our framework considers spatial and channel relation simultaneously. We name our method Dual Relation Distillation (DRD). Unlike \cite{s20164616}, we model spatial and channel relation separately and build two relation maps for both teacher and student. Additionally, we use a different method to build the channel relation map.




 


In summary, we propose a novel Dual Relation Distillation (DRD) framework to transfer both spatial and channel relations in feature maps from the teacher model to the student model. Our proposed method shows better performance compared to previous works in aerial scenes. With a combination of two other useful techniques, our DRD significantly boosts the accuracy of the student model in two aerial benchmarks. We further test the generalization performance of our network in general scenes on the popular Cityscapes dataset. The results show that our DRD achieves better or comparable performance compared to previous works.

\par The rest of this paper is organized as follows. We briefly review some related works  and further clarify the differences with our approach in Section~\ref{sec:rela}. Section~\ref{sec:method} introduces the technical details of the proposed dual relation distillation framework. Section~\ref{sec:experiments} presents the experimental results and conducts the analysis. Finally, Section~\ref{sec:conclusion} draws conclusions.}

\vspace{-0.2cm}
\section{Related Work}
\label{sec:rela}
\vspace{-0.2cm}
In this section, we briefly review some related works on semantic segmentation, and knowledge distillation. 

\subsection{Semantic Segmentation}

{Semantic segmentation is a fundamental topic in computer vision, and the recent rapid development of deep neural networks (DNNs) has had a tremendous impact on its progress. Following the pioneering work \cite{Long2014Fully} that adopts fully convolutional networks for semantic segmentation, many efforts have been made to boost segmentation performance by exploiting multi-scale context. For instance, Chen et al.\cite{Chen2018DeepLab} utilizes dilated convolution to enlarge the receptive field and preserve the spatial size of the feature map. Chen et al. \cite{Chen_2018_ECCV} further develop DeeplabV3+ with an encoder-decoder structure to recover spatial information. PSPNet \cite{zhao2017pspnet} applies pyramid pooling to aggregate contextual information. Hu et al.\cite{hu2021lightweight} present a lightweight and efficient network model LADNet for real-time semantic segmentation tasks. Gao et al. \cite{gao2021mscfnet} also present a novel lightweight network MSCFNet that uses an asymmetric encoder-decoder architecture and efficient attention modules to achieve a good balance between accuracy, inference speed, and model size for real-time semantic segmentation applications. Zhou et al. \cite{ZHOU2022108290} presents a network CENet than employs the encoder-decoder architecture to fully explore contextual features through an ensemble of dense deconvolutions.

Recently, attention mechanisms have been used to guide network learning and alleviate inconsistency in segmentation. For example, \cite{Yu_2018_CVPR} adopts channel attention to select features, while \cite{Fu_2019_CVPR} considers the combination of spatial and channel attention. These state-of-the-art methods aim to boost segmentation performance at the cost of high computational resources.

To address this issue, highly efficient semantic segmentation has been recently studied. 
ESPNet \cite{Mehta_2018_ECCV} designs an efficient spatial pyramid module that decomposes the standard convolution into point-wise convolution followed by spatial pyramid to reduce computational cost. In \cite{Zhao_2018_ECCV}, the authors propose ICNet, an image cascade network based on the compressed PSPNet for real-time semantic segmentation. Yu et al.\cite{Yu_2018_ECCV} introduce BiSeNet, which contains a spatial path and a context path to raise efficiency.}

\subsection{Knowledge Distillation}

{In recent years, knowledge distillation has been extensively researched. 
Knowledge distillation enables the student model to achieve high accuracy while maintaining efficiency. Several knowledge distillation schemes have been proposed recently.
Zagoruyko et al. \cite{Zagoruyko2017AT} transfer the attention map of the teacher model to the student model. Yim et al. \cite{yim2017gift} consider the flow knowledge between layers. \cite{Peng_2019_ICCV} introduce correlation congruence for knowledge distillation to transfer not only the instance-level information but also the correlation between instances. Xu et al. \cite{DBLP:conf/bmvc/0002HH18,PR_1} apply knowledge distillation based on conditional adversarial networks.

Early works of knowledge distillation have primarily focused on image classification. However, with the growing interest in this topic, knowledge distillation approaches have been introduced into other vision tasks, including semantic segmentation. \cite{He_2019_CVPR} adapts knowledge distillation with an additional auto-encoder and transfers densely pairwise affinity maps to the student model. \cite{Liu_2019_CVPR} proposes structured knowledge distillation (SKD), which transfers pairwise relations and forces the outputs of the student model to mimic the teacher model from a holistic view via adversarial learning.\cite{wang2020ifvd} proposes intra-class feature variation distillation (IFVD) to transfer the intra-class feature variation (IFV) of the cumbersome model to the compact model. The self-attention distillation (SAD) is introduced in \cite{Hou_2019_ICCV} to explore attention maps derived from high-level features as the distillation target for shallow layers.

For a given feature map, contextual relationships not only exist in the spatial dimension but also in the channel dimension. However, \cite{Liu_2019_CVPR} only models the relationship in the spatial dimension and does not utilize the information in the channel dimension. Recently, some works have become concerned about channel correlation. \cite{s20164616} models channel and spatial correlation in a single matrix for both teacher and student, then aligns the two matrices using a channel and spatial correlation loss function. \cite{shu2020cwd} applies the softmax normalization to transfer the activation values in each channel into a distribution, then directly minimizes the discrepancy of the channel-wise distribution between the teacher and student network.
However, 
\cite{shu2020cwd} ignores the relation between different channels and implement channel distillation separately for each channel. In contrast, the proposed Dual Relation Distillation (DRD) framework models spatial and channel relations simultaneously. We build two relation maps for both teacher and student, which is different from \cite{s20164616}.} And others also do some meanningful works in segmentation \cite{densenet,uhrsnet,tgrs1,decouple,mbnet,tgrs2,liminglong,zhaoyuzhong,acmmm,boosting1,data,zhaoguiqin,lifelong,cognitive,dlnet,rs2,fusing,ldnet,organizing,edge,binary,energy,boosting,synthetic,dynrsl,flexdataset,gmm,llmcot,geogrambench,geolocsft,f2net}.



\section{Method}
\label{sec:method}
\begin{figure*}[h]
  \centering
  \includegraphics[width=1.00\linewidth]{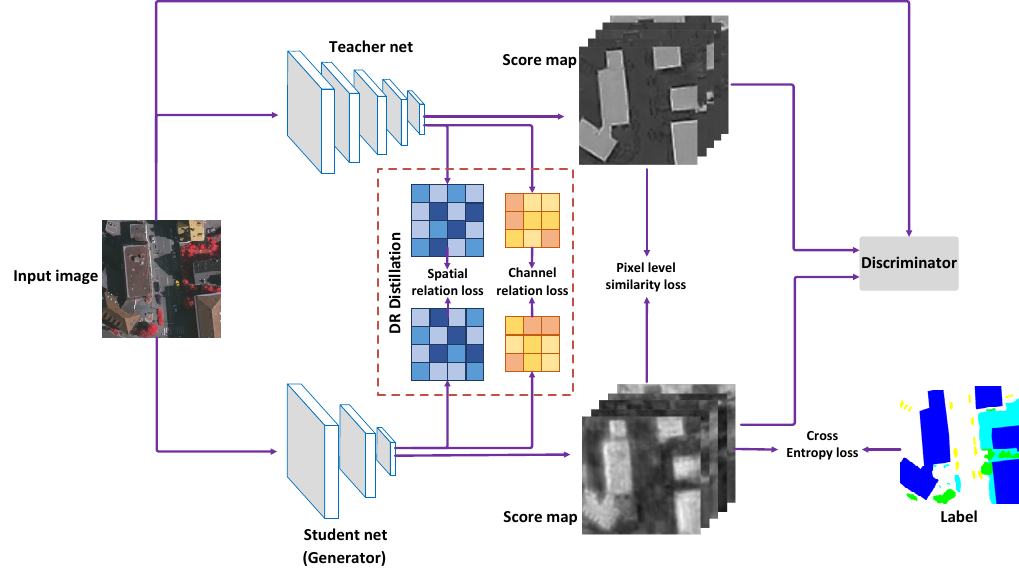}
  \caption{Pipeline of the proposed Dual Relation Distillation (DRD). (Example from Vaihingen dataset, best viewed in color.)}
  \label{pic:mainfig}
\end{figure*}

{In this section, we present the proposed Dual Relation Distillation (DRD) pipeline. An overview of our method is illustrated in Figure ~\ref{pic:mainfig}. We design the teacher network to have a deeper encoder than the student network. To effectively transfer knowledge from the teacher network to the student network, we equip our pipeline with four types of loss functions, in addition to the commonly used cross-entropy loss for pixel categorization.}
\vspace{-0.2cm}
\subsection{Dual Relation Alignment in Feature Map}

{Due to the different sizes of the receptive field in the student and teacher networks, the features corresponding to pixels with the same label may have some differences. Therefore, we propose a spatial relation loss and a channel relation loss to transfer rich information from the teacher feature map to the student feature map.

Given an input feature map $\mathcal{F} \in \mathbb{R}^{C \times H \times W}$, where $W$ and $H$ are the width and height of the feature map, respectively, and $C$ is the number of channels, we first reshape it to matrix $\mathbf{F} \in \mathbb{R}^{C \times N}$, where $N=H \times W$ is the total number of pixels in the feature map. Next, we perform a matrix multiplication $ \mathbf{F}^{T} \mathbf{F} $ between $\mathbf{F}$ and its transpose, $\mathbf{F}^{T}$. We then apply a softmax layer to calculate the spatial relation map, $\mathbf{S} \in \mathbb{R}^{N \times N}$,

\begin{equation}
s_{i j}=\frac{\exp \left(F_{j}^{T} F_{i}\right)}{\sum_{j=1}^{N} \exp \left(F_{j}^{T} F_{i}\right)}
\end{equation}
where $ F_{i} $ denotes the feature of pixel $i$, which corresponds to the  $i$th column of matrx $\mathbf{F}$, and $s_{i j}$ denotes the normalized affinity of pixel $j$ with respect to pixel $i$ .  The larger the value $s_{i j}$ in the relation map  $\mathbf{S}$ is, the stronger the relative correlation  of pixel $j$ with respect to pixel $i$ is.

To minimize the distance between the spatial relation maps of the teacher network $t$ and the student network $s$, we utilize the conventional Mean Squared (L2) loss as shown below:
\begin{equation}
L_{S}=\frac{1}{\left(N \times N\right)} \sum_{i=1}^{N} \sum_{j=1}^{N}\left(s_{i j}^{s}-s_{i j}^{t}\right)^{2} .
\end{equation}


Spatial relations mainly capture long-range dependencies from the teacher network, which are difficult to learn for small models due to their limited receptive field and abstraction capability. 
For a given feature map, contextual relationships not only exist in the spatial dimension but also in the channel dimension, especially in high-level features. Intuitively, each channel of the high-level feature map can be regarded as a class-specific response, and different semantic responses are associated with each other.
Therefore, we can evaluate a channel relation map $\mathbf{C}=(c_{i j}) \in \mathbb{R}^{C \times C}$, and we propose a channel relation loss similar to the spatial relation loss for training the student network. The loss function is defined as follows:

\begin{equation}
L_{C}=\frac{1}{\left(C \times C\right)} \sum_{i=1}^{C} \sum_{j=1}^{C}\left(c_{i j}^{s}-c_{i j}^{t}\right)^{2} ,
\end{equation}
where $c_{i j}$ denotes the normalized affinity of channel $j$ with respect to channel $i$, and the superscripts $s$ and $t$ is used to distinguish between the student and teacher networks.  The normalized affinity $c_{i j}$ is computed by 
\begin{equation} 
c_{i j}=\frac{\exp \left(\hat{F}_{j} \hat{F}_{i}^{T}\right)}{\sum_{j=1}^{C} \exp \left(\hat{F}_{j} \hat{F}_{i}^{T}\right)},
\end{equation}
where $\hat{F}_{i} $ denotes the features of all pixels in channel $i$, which corresponds to the $i$th row of
matrx $\mathbf{F}$, and $c_{i j}$ denotes the normalized affinity of channel $j$ with respect to channel $i$ . The larger the value $c_{i j}$ in the relation map $\mathbf{C}$ is, the stronger the relative
correlation of channel $j$ with respect to channel $i$ is.




Considering that the student model should not only mimic the feature distribution, but also the output score map of the teacher model, we also include the loss of original knowledge distillation and adversarial learning in our pipeline.}

\subsection{Pixel-Level Similarity Alignment in Score Map}

{We view the segmentation problem as a collection of individual pixel labeling tasks and utilize knowledge distillation to align the class probabilities of each pixel generated by the compact network. Our approach is straightforward: we use the class probabilities generated by the cumbersome model as soft targets to train the compact network.

The loss function is defined as follows,
\begin{equation}
L_{P}=\frac{1}{N} \sum_{i=1}^{N} \mathrm{KL}\left(\mathbf{q}_{i}^{t} \| \mathbf{q}_{i}^{s}\right)
\end{equation}
where $\mathbf{q}_{i}^{s}$ represents the class probabilities of the $i$ th pixel produced from the compact network s, and $\mathbf{q}_{i}^{t}$ represents the class probabilities of the $i$th pixel produced from the cumbersome network t.
Both $\mathbf{q}_{i}^{s}$ and $\mathbf{q}_{i}^{t}$ are vectors of size  $1 \times 1 \times c$, where $c$ denotes the number of classes in the segmentation task. $N$ denotes the total number of pixels in the feature map. $\mathrm{KL}(\cdot)$ denotes the Kullback-Leibler divergence between two probabilities, which is calculated as follows:

\begin{equation}
\operatorname{KL}\left(\mathbf{q}_{i}^{t} \| \mathbf{q}_{i}^{s}\right)=\sum_{i=1}^{N} \sum_{j=1}^{c} \mathbf{q}_{i j}^{t} \cdot \log \frac{\mathbf{q}_{i j}^{t}}{\mathbf{q}_{i j}^{s}},
\end{equation}
where $\mathbf{q}_{i j}^{t}$ and $\mathbf{q}_{i j}^{s}$ denote the $j$th entry of class probability vectors $\mathbf{q}_{i}^{s}$ and $\mathbf{q}_{i}^{t}$, respectively.}

\subsection{Global Similarity Alignment through Adversarial Learning}

{We employ an adversarial training approach to ensure that the holistic embeddings of the segmentation maps generated by the compact segmentation network are indistinguishable from those produced by the cumbersome segmentation network. This technique, known as adversarial learning for knowledge distillation, was first introduced in \cite{DBLP:conf/bmvc/0002HH18}. A similar approach, called holistic distillation, was proposed in \cite{Liu_2019_CVPR} for semantic segmentation. We also incorporate adversarial learning in the output space. Specifically, we first train a discriminator to differentiate between inputs from the teacher model and the student model by evaluating the similarity between the raw image and the segmentation map. The segmentation network is then trained to deceive the discriminator. The loss functions for training the discriminator ($L_{d}$) and the adversarial item ($L_{adv}$) can be formulated as follows:
\begin{equation}
L_{d}=\mathbb{E}_{z_{s} \sim p_{s}\left(z_{s}\right)}\left[D\left(z_{s} \mid I\right)\right]-\mathbb{E}_{z_{t} \sim p_{t}\left(z_{t}\right)}\left[D\left(z_{t} \mid I\right)\right],
\end{equation}

\begin{equation}
L_{A d v}=\mathbb{E}_{z_{s} \sim p_{s}\left(z_{s}\right)}\left[D\left(z_{s} \mid I\right)\right].
\end{equation}

For our proposed dual relation distillation (DRD), the total loss as the training objective function is composed of a conventional cross-entropy loss $L_{C E}$ for semantic segmentation and three loss items for knowledge distillation:
\begin{equation}
\operatorname{L_{total}} =L_{C E}+\lambda_{1} L_{P}-\lambda_{2} L_{A d v}+\lambda_{3} \left(L_{S} + L_{C}\right),
\end{equation}
where $\lambda_{1}, \lambda_{2}, \lambda_{3}$ are set to 10, 0.1 and 25, respectively.}

\section{Experiments}
\label{sec:experiments}


\subsection{Datasets}
\subsubsection{Vaihingen Dataset}

{The Vaihingen dataset comprises 33 aerial images covering a 1.38 $\mathrm{km}^{2}$ area of the city of Vaihingen, with a spatial resolution of 9 cm. Each image has an average size of 2494 $\times$ 2064 pixels and three bands corresponding to near-infrared (NIR), red (R), and green (G) wavelengths. Additionally, DSMs, which indicate the height of all object surfaces in an image, are provided as complementary data. Out of the 33 images, 16 are manually annotated with pixel-wise labels, and each pixel is classified into one of six land cover classes: buildings, cars, low vegetation, impervious surfaces, trees, and clutter/background. Clutter/background is not considered when counting loss. Following the setup in \cite{Mou_2019_CVPR}, we use 11 images for training, and the remaining five images (image IDs: 11, 15, 28, 30, 34) are used to test our model.}

\subsubsection{Potsdam Dataset} 

{The Potsdam dataset comprises 38 high-resolution aerial images covering an area of 3.42 $\mathrm{km}^{2}$, with each image captured in four channels (NIR, R, G, and blue (B)). All images are of size 6000 $\times$ 6000 pixels and are annotated with pixel-level labels of six classes, similar to the Vaihingen dataset. The spatial resolution is 5 cm, and coregistered DSMs are also available. For training and evaluating networks, we use 17 images for training and build the test set with the remaining images (image IDs: 02\_11, 02\_12, 04\_10, 05\_11, 06\_07, 07\_08, 07\_10), following the setup in \cite{Mou_2019_CVPR}.}

\subsubsection{Cityscapes Dataset}

The Cityscapes dataset is designed for urban scene understanding and consists of 30 classes, with only 19 classes used for evaluation. The dataset includes 5,000 high-quality pixel-level finely annotated images and 20,000 coarsely annotated images. The finely annotated 5,000 images are divided into 2,975/500/1,525 images for training, validation, and testing. In our experiments, we only use the finely annotated parts of the dataset in our experiments.

\subsection{Evaluation Metric}

To evaluate the performance of our DRD, we introduce various evaluation metrics. Firstly, we calculate $F_{1}$ score for each category with the following formula:
\begin{equation} 
F_{1}=\left(1+\beta^{2}\right) \cdot \frac{\text {precision } \cdot \text {recall}}{\beta^{2} \cdot \text {precision }+\text {recall}}, \quad \beta=1.
\end{equation}
Furthermore,  we compute the mean $F_{1}$ score by averaging all $F_{1}$ scores to assess models impartially. Notably, a higher $F_{1}$ score indicates a better result. 

In addition, we also calculate the Overall Accuracy (OA) and mean Intersection over Union (mIOU) of all categories to provide a detailed and comprehensive comparison with different methods. The Overall Accuracy is calculated by counting the proportion of correctly classified pixels to all pixels. The concrete computation is defined as follows,

\begin{equation}
\mathrm{OA}=\frac{\sum_{i=1}^{M} p_{i i}}{\sum_{i=1}^{M} \sum_{j=1}^{M} p_{i j}}
\end{equation}
where $p_{i j}$ denotes the number of pixels that actually belong to category $i$ and are predicted to be category $j$, and $M$ is the total number of all the categories.

For a specific category, the intersection and union ratio can be calculated using the following formula:

\begin{equation}
\mathrm{IoU}=\frac{|y \cap \hat{y}|}{|y \cup \hat{y}|},
\end{equation}
where $\hat{y}$ and $y$ represent the region corresponding to the category in the predicted segmentation result and the ground truth, respectively. The accuracy of the model's prediction is measured by the overlap ratio of the two regions. The mean intersection over union ratio is the unweighted average of the intersection over union ratio of each category.

Additionally, we also calculate the model size and complexity. The model size is represented by the number of network parameters, while the model complexity is evaluated by the sum of floating point operations (FLOPs) in one forward pass on a fixed resolution of 512 × 1024 using the PyTorch implementation \footnote{https://github.com/warmspringwinds/pytorch-segmentation-detection/blob/master/pytorch\_segmentation\_detection/utils/flops\_benchmark.py}

\subsection{Implementation details}

\subsubsection{Network architectures}

{To ensure a fair comparison, we conduct the experiments using the same cumbersome and compact networks as \cite{Liu_2019_CVPR} and \cite{wang2020ifvd}. Specifically, we use the segmentation architecture PSPNet \cite{zhao2017pspnet} with ResNet101 \cite{He_2016_CVPR} backbone as the teacher model for all experiments. The student model also utilizes PSPNet as the segmentation architecture, but with different backbone networks. We conduct the experiments on ResNet18 \cite{He_2016_CVPR} and ResNet18 (0.5), which is the width-halved version of ResNet18, as the backbone for the student model.}

\subsubsection{Training details}

For aerial benchmark, since the resolution of remote sensing image is too high, it cannot be directly used for model training and testing. When training with the subsampled high-resolution remote sensing images, the segmentation performance is poor thanks to the loss of too much detailed information. Therefore, we choose to crop the high-resolution remote sensing image into small slices for training and testing. We choose image crop size as $645 \times 645$ for Vaihingen dataset and $600 \times 600$ for Potsdam dataset. As for Cityscapes dataset we do not crop the initial images. The implementation is based on the PyTorch platform. All our distillation experiments are carried out on a workstation equipped with a single NIVIDIA RTX 3090 GPU card of $24 \mathrm{~GB}$ memory.

\subsection{Ablation Study}

\begin{table}[H]
\centering
  \caption{Ablation study on Vaihingen dataset. ImN means initial from the ImageNet-pre-trained weights}
  \label{res:ablation}
  \scalebox{0.9}{\begin{tabular}{l|cccc|c|c|c}
    \toprule
     Method & $L_{P}$ & $L_{Adv}$ & $L_{S}$ & $L_{C}$ & mean $F_{1}$(\%) & mIoU(\%) & OA(\%)  \\ 
    \midrule
    Teacher(ResNet101) &  &  & & & 86.07 & 75.99 & 86.67 \\
    \midrule
    ResNet18 & &  & & &79.15 & 66.53 & 83.09 \\

    ResNet18 & \checkmark &  & & &79.88 & 67.47 & 83.73\\
    ResNet18 & \checkmark & \checkmark & & &80.16 & 67.79 & 83.79\\
    ResNet18 & \checkmark & \checkmark & \checkmark & &80.40 & 68.18 & 84.18\\
    ResNet18 & \checkmark & \checkmark & \checkmark & \checkmark & 80.83 & 68.62 & 83.94 \\
    \midrule
    ResNet18(IMN) & &  & & & 81.13 & 68.94 & 83.74 \\
    ResNet18(IMN) & \checkmark &  & & & 82.81& 71.27& 84.89 \\
    ResNet18(IMN) & \checkmark & \checkmark & & & 83.12& 71.69& 85.27\\
    ResNet18(IMN) & \checkmark & \checkmark & \checkmark & & 83.59 & 72.33 & 85.26\\
    ResNet18(IMN) & \checkmark & \checkmark & \checkmark & \checkmark & 84.03 & 72.95 & 85.46 \\
    \bottomrule
\end{tabular}
}
\end{table}

{In this section, we conduct ablation studies on the Vaihingen dataset to demonstrate the effectiveness of our technical contributions in Section~\ref{sec:method}. Our proposed DRD includes four loss items for knowledge distillation: $L_{P}$, $L_{Adv}$, $L_{S}$, and $L_{C}$. Therefore, we study their contributions individually. We first train the teacher model PSPNet with ResNet101 backbone and then perform knowledge distillation on the student model with ResNet18 backbone. Table~\ref{res:ablation} presents the results of different settings for the student network. As shown, knowledge distillation improves the performance of the student network, and distilling the relation information in the feature map helps the student learn better. With the four distillation terms, the improvements for ResNet18 and ResNet18 with ImageNet-pretrained weights are $1.68\%, 2.90\%$ in mean $F_{1}$ and $2.10\%, 4.01\%$ in mIoU, respectively. Additionally, each distillation scheme leads to a higher mean $F_{1}$ score and mIoU score, indicating that the four distillation schemes make complementary contributions to better train the compact network. After distillation, the best mIoU for ResNet18(IMN) reaches $72.95\%$ on the test set, and the gap between the student and teacher is reduced from $7.05\%$ to $3.04\%$ when both are initialized from ImageNet-pretrained weights. These results demonstrate the effectiveness of the proposed DRD.}

\subsection{Comparison with Existing Works}

\begin{table}[H]
\centering
  \caption{Experimental Results on the Vaihingen Dataset. }
  \label{res:vaihingen}
  \scalebox{0.57}{  \begin{tabular}{l|ccccc|c|c|c|c|c}
    \toprule
    Method &  Imp. surf. & Build. & Low veg. & Tree & Car & mean $F_{1}$ & mIoU & OA & Params(M) & FLOPs(G) \\
    \midrule
    \midrule
    FCN-dCRF (\cite{journals/corr/ChenPKMY14}) & 88.80 & 92.99 & 76.58 & 86.78 & 71.75 & 83.38 & 72.28 & 86.65 & - & - \\ 
    FCN (\cite{Long2014Fully}) & 88.67 & 92.83 & 76.32 & 86.67 & 74.21 & 83.74 & 72.69 & 86.51  & 134.5 & 333.9\\
    SCNN (\cite{scnn_}) & 88.21 & 91.80 & 77.17 & 87.23 & 78.60 & 84.40 & 73.73 & 86.43 & - & - \\
    Dilated FCN (\cite{Chen2018DeepLab}) & 90.19 & 94.49 & 77.69 & 87.24 & 76.77 & 85.28 & - & 87.70 & 262.1 & 457.8\\
    FCN-FR (\cite{MaggioriHigh}) & 91.69 & 95.24 & 79.44 & 88.12 & 78.42 & 86.58 & - & 88.92 & - & - \\
    S-RA-FCN \cite{Mou_2019_CVPR} & 91.47 & 94.97 & 80.63 & 88.57 & 87.05 & 88.54 & 79.76 & 89.23 & - & - \\
    \midrule
    Teacher: PSPNet(ResNet101) & 89.38  & 94.19  & 76.80  & 85.40  &  84.57 &  86.07 & 75.99 & 86.67  & 70.43 & 574.9  \\
    \midrule
    ResNet18(0.5) &  85.15 & 90.46  & 72.24  &  82.57 & 61.48  & 78.38  & 65.59  & 82.69 & 3.27 & 31.53  \\
    ResNet18(0.5) + IFVD* \cite{wang2020ifvd} & 84.70  &  90.37 & 72.09  & 82.73  & 65.31 &  
    79.04 & 66.26 & 82.59  & 3.27 & 31.53\\
    ResNet18(0.5) + SKD* \cite{Liu_2019_CVPR} & 86.60  &  91.11 & 73.65  & 83.33  & 67.93 & 80.52 & 68.24 & 83.82 & 3.27 & 31.53\\  
    ResNet18(0.5) + DRD & 86.69  &  90.96 & 73.73  & 83.42  & 71.77 & 81.31  & 69.17  & 83.91 & 3.27 & 31.53\\   
    \midrule
    ResNet18 & 86.30 & 91.26  & 73.19  & 83.48  & 71.39  & 81.13  & 68.94  & 83.74 & 13.07 & 125.8  \\
    ResNet18 + IFVD* \cite{wang2020ifvd} & 87.04  & 92.46  &  74.06 & 84.08  & 75.91  
    &  82.71 & 71.11 & 84.58  & 13.07 & 125.8\\
    ResNet18 + SKD* \cite{Liu_2019_CVPR} & 87.84  & 92.62  &  75.44 & 84.51  & 77.54  &  83.59 & 72.33 & 85.26  & 13.07 & 125.8\\ 
    ResNet18 + DRD & 87.99  & 92.64  & 75.28  & 84.90  & 79.36 &  84.03 & 72.95  & 85.46 & 13.07 & 125.8\\
  \bottomrule
\end{tabular}}
\end{table}

\begin{figure}[H]
  \centering
  \includegraphics[width=1.0\linewidth]{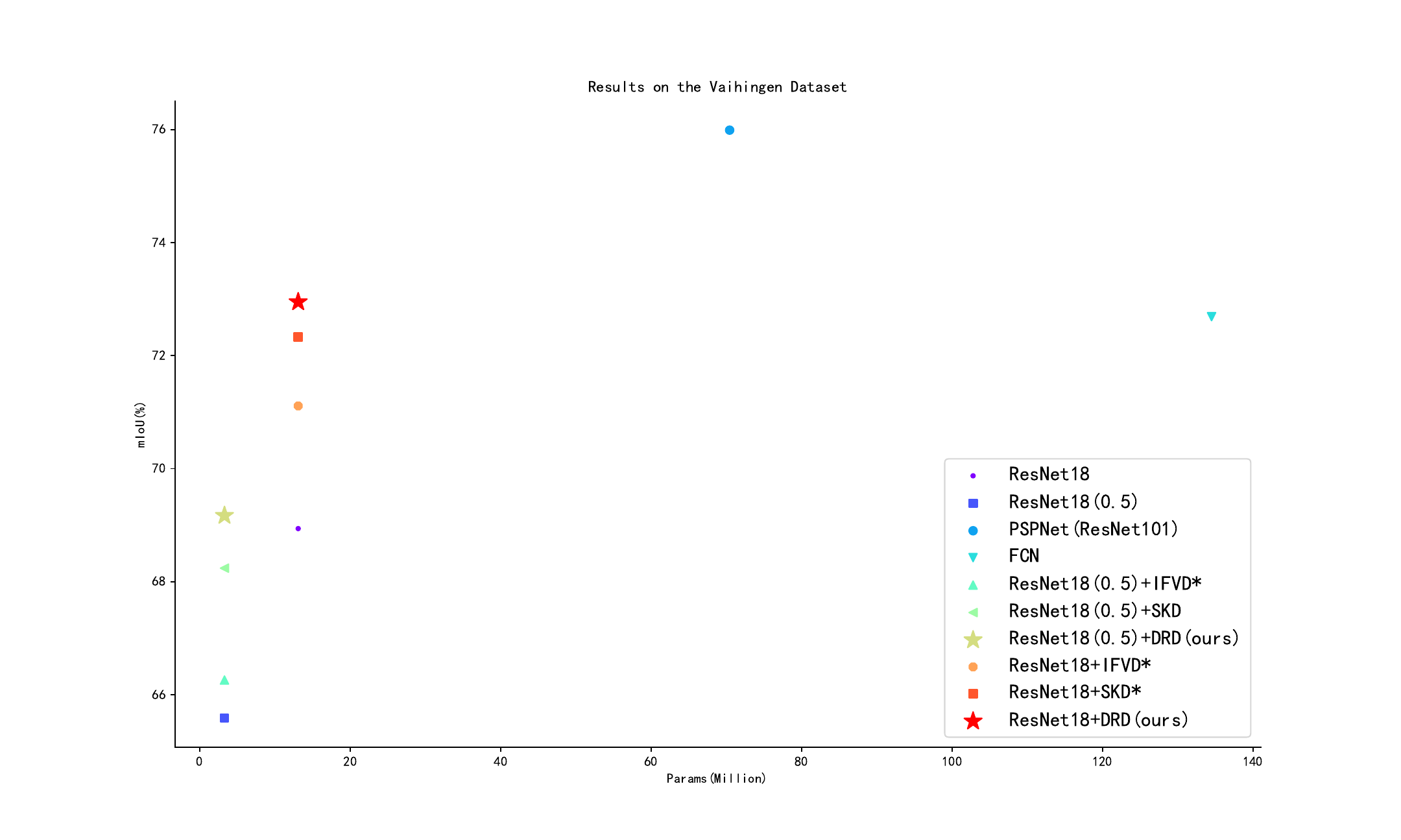}
  \caption{Performance (mIoU (\%)) vs. models sizes of some methods in Table~\ref{res:vaihingen}  on Vaihingen Dataset.}
  \label{pic_vaihingen}
\end{figure}

{We then evaluate our proposed DRD on the Vaihingen dataset and compare it with existing works. The quantitative results are listed in Table~\ref{res:vaihingen}, which includes the F1 score of each class, overall segmentation accuracy (mean $F_{1}$, mIoU, OA), model size, and model complexity. Since the methods of \cite{Liu_2019_CVPR} and \cite{wang2020ifvd} do not provide experimental results in aerial datasets, we evaluate \cite{Liu_2019_CVPR} and \cite{wang2020ifvd} with their given codes in the Vaihingen dataset, and the results are denoted by *.}

{Our distillation approach has shown significant improvements in three evaluation metrics for two compact networks: ResNet18 (0.5) and ResNet18. Specifically, our DRD improves the student model built on ResNet18 without distillation by $2.90 \%, 4.01 \%$, and $1.72 \%$ in mean $F_{1}$, mIoU, and Overall Accuracy, respectively, with a gain of $0.44 \%, 0.62 \%$, and $0.20 \%$ compared to the previous method \cite{Liu_2019_CVPR}. We also applied the proposed distillation scheme on another backbone, ResNet18 (0.5), a width-halved version of the original ResNet18 without ImageNet-pre-trained weights. The proposed DRD improves the student model based on ResNet18 (0.5) backbone by $2.93 \%, 3.58 \%$, and $1.22 \%$ in mean $F_{1}$, mIoU, and Overall Accuracy, respectively, with a gain of $0.79 \%, 0.93 \%$, and $0.09 \%$ compared to \cite{Liu_2019_CVPR}. These results demonstrate the effectiveness of the proposed DRD method. In order to more clearly demonstrate the performance and model size correlation between the pre and post distillation models, as well as the compact and cumbersome models, we have visually summarized some methods in Table~\ref{res:vaihingen} in Figure ~\ref{pic_vaihingen}

\begin{figure}[H]
  \centering
  \includegraphics[width=\linewidth]{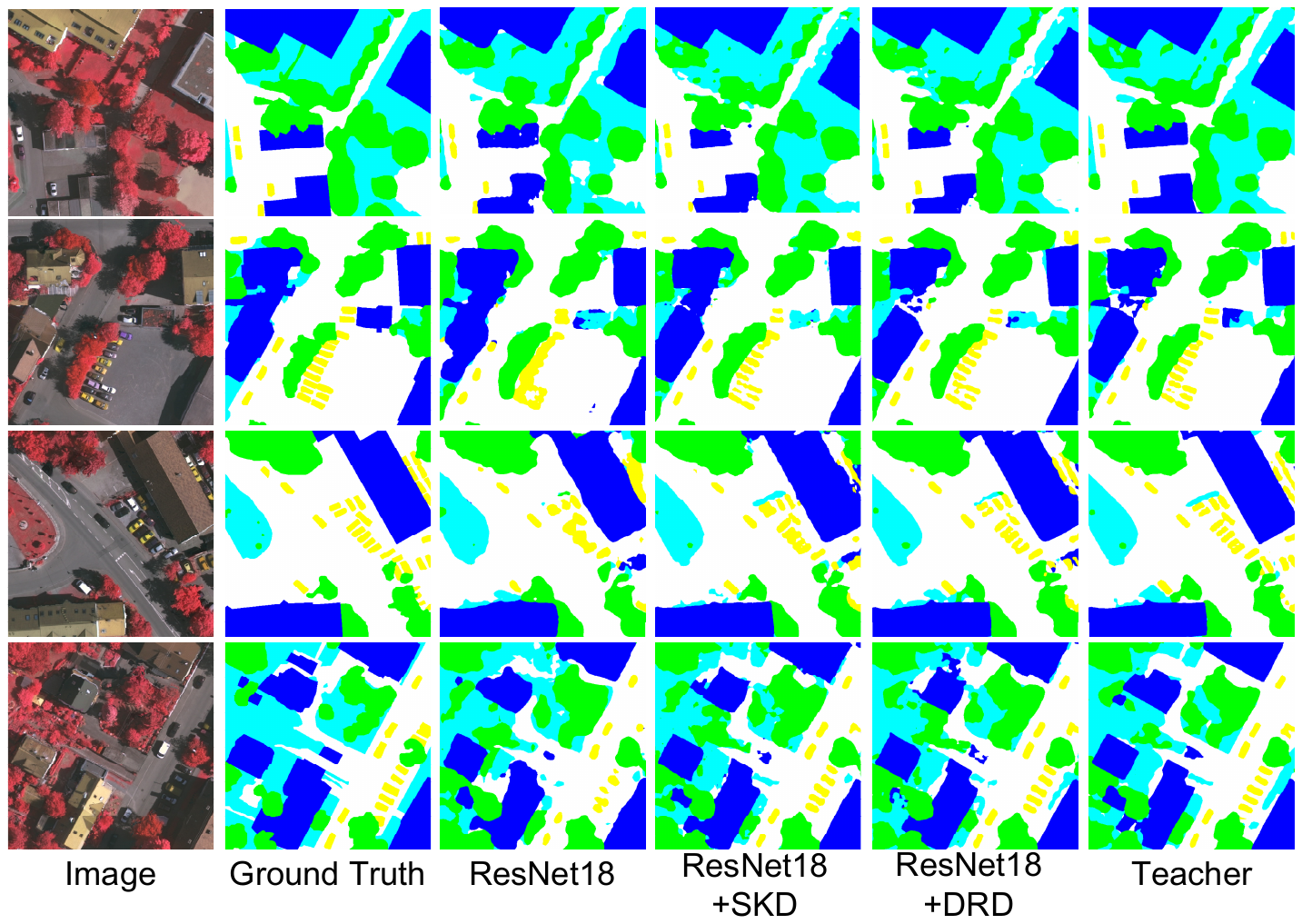}
  \caption{Examples of segmentation results on the Vaihingen dataset. Legend—white: impervious surfaces, \textcolor{blue}{blue}: buildings, \textcolor{cyan}{cyan}: low vegetation, \textcolor{green}{green}: trees, \textcolor{yellow}{yellow}: cars, \textcolor{red}{red}: clutter/background. (Best viewed in color.)}
  \label{pic:vaihingenres}
\end{figure}

Figure~\ref{pic:vaihingenres} shows a few examples of qualitative results on the Vaihingen dataset. By comparing column 3 and column 5, we can clearly see that after distillation using our DRD framework, the student network generates more accurate segmentation results compared to no extra guidance from the teacher network, especially for the previous low-accuracy class such as cars. By comparing column 4 and column 5, we can see that our DRD leads to better results compared to \cite{Liu_2019_CVPR}. If we compare the last column with column 3 and column 5, we can conclude that the gap between the student model and teacher model is reduced.

\begin{figure}[H]
  \centering
  \includegraphics[width=\linewidth]{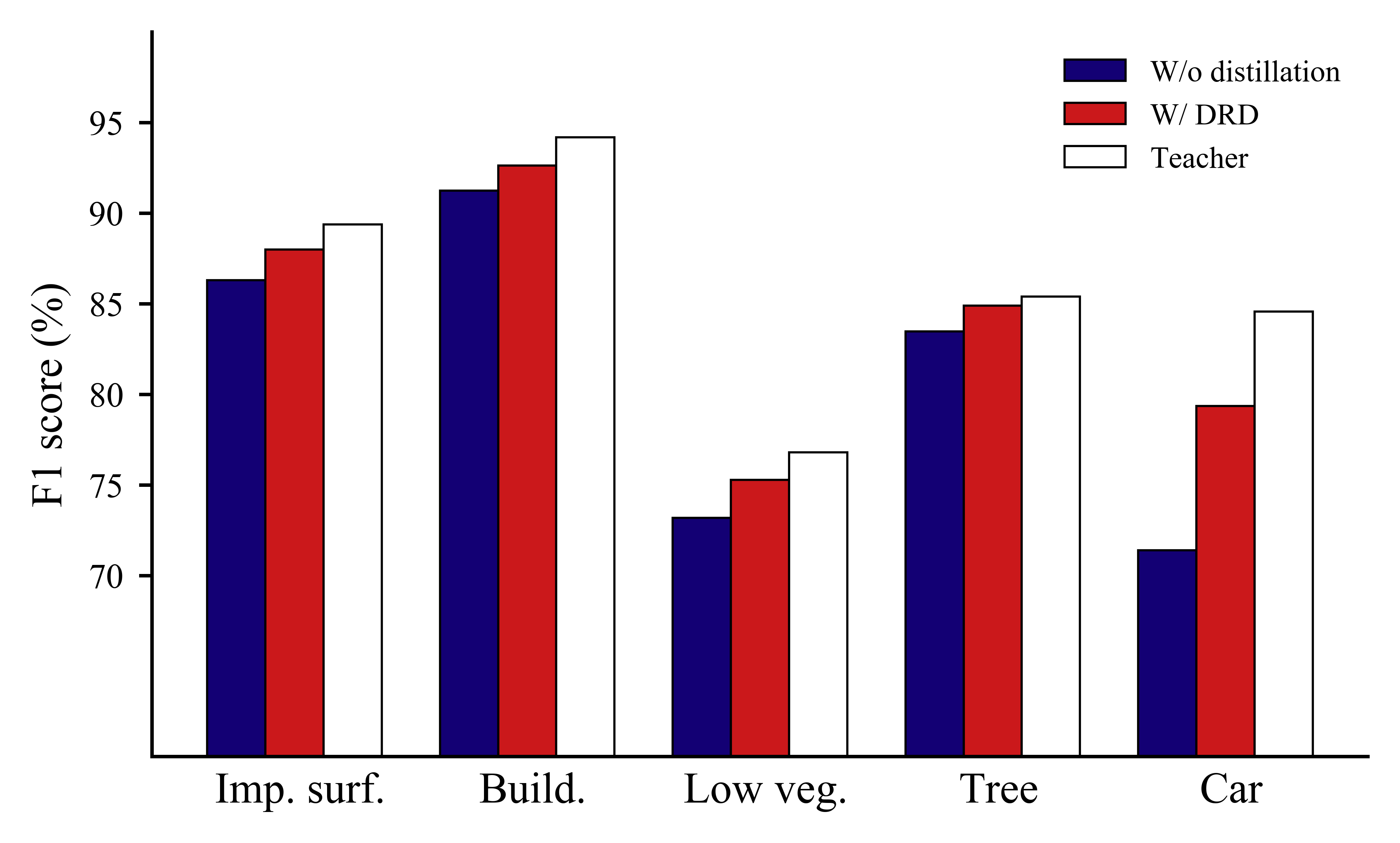}
  \caption{Effectiveness of our DRD on the Vaihingen dataset with ResNet 18.}
  \label{pic:vaihingenzhu}
\end{figure}
Figure~\ref{pic:vaihingenzhu} shows the $F_{1}$ score improvements for each category in the student network using ResNet(18) backbone after distillation. As can be seen, our distillation method improves all categories, especially for categories with low accuracy, and reduces the gap between the teacher model and student model after distillation.}

\subsection{Results on the Potsdam Dataset}

\begin{table}[H]
\centering
  \caption{Experimental Results on the Potsdam Dataset.}
  \label{res:Potsdam}
  \scalebox{0.57}{  \begin{tabular}{l|ccccc|c|c|c|c|c}
    \toprule
    Method &  Imp. surf. & Build. & Low veg. & Tree & Car & mean $F_{1}$ & mIoU & OA & Params(M) & FLOPs(G) \\
    \midrule
    \midrule
    Dilated FCN \cite{Chen2018DeepLab} & 86.52 & 90.78 & 83.01 & 78.41 & 90.42 & 85.83 & - & 84.14 & 262.1 & 457.8 \\
    SCNN \cite{scnn_} & 88.37 & 92.32 & 83.68 & 80.94 & 91.17 & 87.30 & 77.72 & 85.57 & - & - \\
    FCN \cite{Long2014Fully} & 88.61 & 93.29 & 83.29 & 79.83 & 93.02 & 87.61 & 78.34 & 85.59 & 134.5 & 333.9 \\
    FCN-dCRF \cite{journals/corr/ChenPKMY14} & 88.62 & 93.29 & 83.29 & 79.83 & 93.03 & 87.61 & 78.35 & 85.60 & - & - \\ 
    FCN-FR \cite{MaggioriHigh} & 89.31 & 94.37 & 84.83 & 81.10 & 93.56 & 88.63 & - & 87.02 & - & - \\
    S-RA-FCN \cite{Mou_2019_CVPR} & 91.33 & 94.70 & 86.81 & 83.47 & 94.52 & 90.17 & 82.38 & 88.59 & - & - \\
    \midrule
    Teacher: PSPNet(ResNet101) & 90.91  & 94.26  & 86.53  & 82.06  & 91.41  & 89.04  & 80.50 & 88.62 & 70.43 & 574.9 \\
    \midrule
    ResNet18(0.5) & 86.50  & 90.12  & 82.15  & 73.93  & 85.18  & 83.58  & 72.16  & 83.56 & 3.27 & 31.53 \\
    ResNet18(0.5) + SKD* \cite{Liu_2019_CVPR}  & 88.70  & 92.71  & 84.16  & 79.31  & 88.52 & 86.68  & 76.78  & 86.38 & 3.27 & 31.53 \\  %
    ResNet18(0.5) + IFVD* \cite{wang2020ifvd} &  88.94 & 92.78 & 84.22 & 78.27 & 89.44 & 86.73 & 76.91  &  86.28  & 3.27 & 31.53 \\
    ResNet18(0.5) + DRD &  89.29 & 92.89  & 84.83  & 79.85  & 89.68 & 87.31  & 77.76  & 86.92 & 3.27 & 31.53 \\  
    \midrule
    ResNet18 & 88.46  & 92.10  & 83.48  & 77.33  & 88.58  & 85.99  &  75.77 & 85.63 & 13.07 & 125.8 \\
    ResNet18 + SKD* \cite{Liu_2019_CVPR}  & 90.00  & 92.93  & 85.41 & 79.50  & 90.14 & 87.60  & 78.24  & 87.30 & 13.07 & 125.8 \\
    ResNet18 + IFVD* \cite{wang2020ifvd} & 89.66  & 93.24  &  84.99 & 80.65  & 90.07  &  87.72 & 78.40 & 87.23  & 13.07 & 125.8 \\
    ResNet18 + DRD & 90.10  & 93.33  & 85.72  & 81.05  & 90.54 & 88.15  & 79.07  & 87.76 & 13.07 & 125.8 \\
  \bottomrule
\end{tabular}}
\end{table}

\begin{figure}[H]
  \centering
  \includegraphics[width=1.00\linewidth]{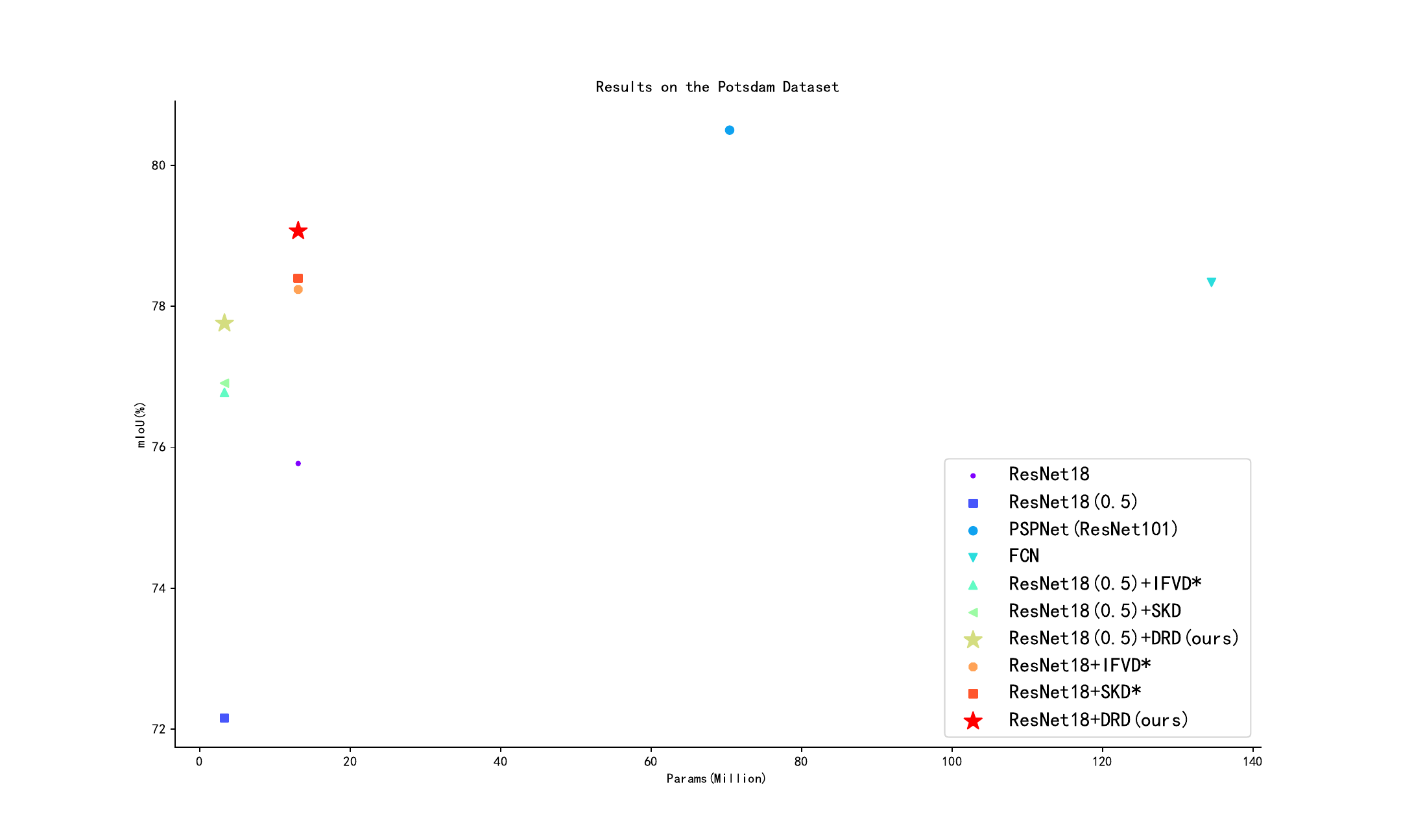}
  \caption{Performance (mIoU (\%)) vs. models sizes of some methods in Table~\ref{res:Potsdam}  on Potsdam Dataset.}
  \label{pic_potsdam}
\end{figure}

{To further validate the effectiveness of our network, we conduct experiments on the Potsdam dataset, and the numerical results are shown in Table~\ref{res:Potsdam}. * denotes the results we reproduce using the released code of \cite{Liu_2019_CVPR} and \cite{wang2020ifvd}. As can be seen, after distillation, our DRD improves the student model built on ResNet18 without distillation by $2.16 \%$, $3.30 \%$, and $2.13 \%$ in mean $F_{1}$, mIoU, and Overall Accuracy, respectively, with a gain of $0.55 \%$, $0.83 \%$, and $0.46 \%$ compared to the previous method \cite{Liu_2019_CVPR}. We also applied the proposed distillation scheme on ResNet18 (0.5), a width-halved version of the original ResNet18 without ImageNet-pre-trained weights. Our proposed method improves the ResNet18 (0.5) model by $3.73 \%$, $5.60 \%$, and $3.36 \%$ in mean $F_{1}$, mIoU, and Overall Accuracy, respectively, with a gain of $0.63 \%$, $0.98 \%$, and $0.54 \%$ compared to \cite{Liu_2019_CVPR}. These results demonstrate the effectiveness of the proposed DRD. After distillation, the gap between the teacher model and student model in the Potsdam dataset is less than the result in the Vaihingen dataset, which is probably due to the availability of more training data. To more clearly demonstrate the performance and model size correlation between the pre and post distillation models, as well as the compact and cumbersome models, we have visually summarized some methods in Table~\ref{res:Potsdam} in Figure ~\ref{pic_potsdam}


\begin{figure}[H]
  \centering
  \includegraphics[width=\linewidth]{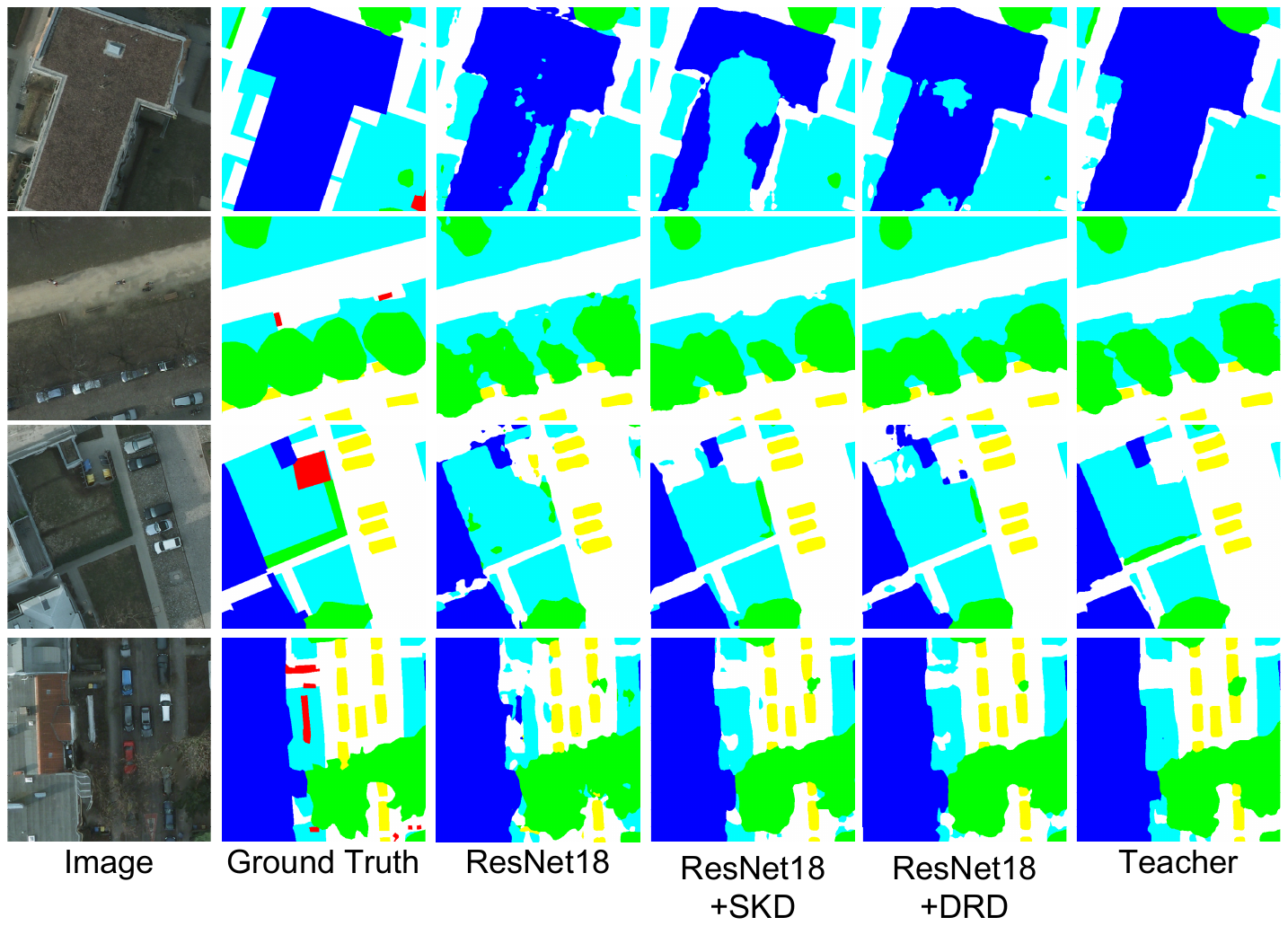}
  \caption{Examples of segmentation results on the Potsdam dataset. Legend—white: impervious surfaces, \textcolor{blue}{blue}: buildings, \textcolor{cyan}{cyan}: low vegetation, \textcolor{green}{green}: trees, \textcolor{yellow}{yellow}: cars, \textcolor{red}{red}: clutter/background. (Best viewed in color.)}
  \label{pic:Potsdamres}
\end{figure}

Figure~\ref{pic:Potsdamres} shows a few examples of qualitative results on the Potsdam dataset. By comparing column 4 and column 5, we can see that our DRD leads to better results compared to \cite{Liu_2019_CVPR}. By comparing column 3 and column 5, we can see that after distillation using our DRD framework, the student network generates more accurate segmentation results compared to no extra guidance from the teacher network, especially for the previous low-accuracy class such as trees. If we compare the last column with column 3 and column 5, we can conclude that the gap between the student model and teacher model is reduced after distillation.


\begin{figure}[H]
  \centering
  \includegraphics[width=\linewidth]{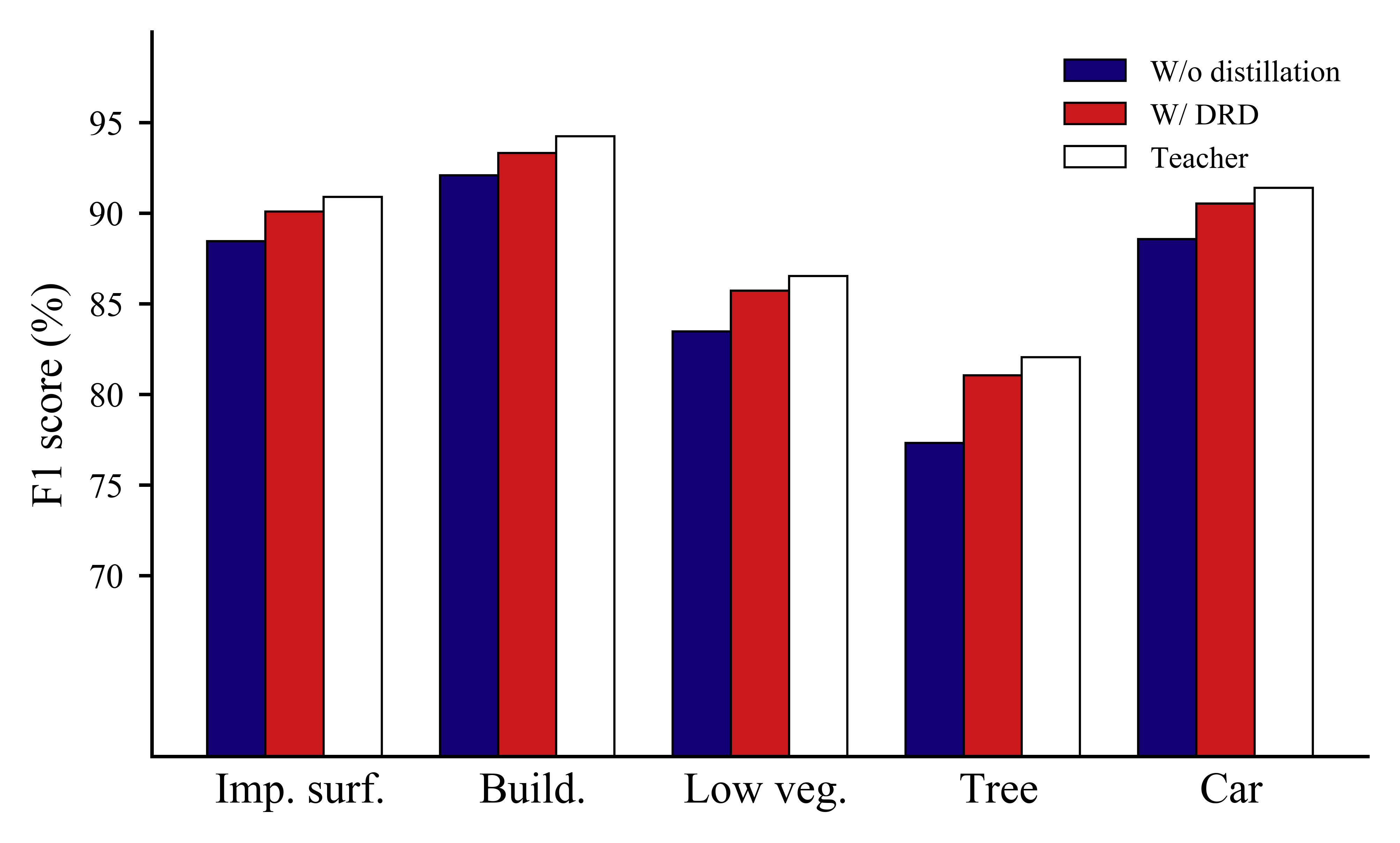}
  \caption{Effectiveness of our DRD on the Potsdam dataset with ResNet 18.}
  \label{pic:Potsdamzhu}
\end{figure}
Figure~\ref{pic:Potsdamzhu} shows the $F_{1}$ score improvements for each category in the student network using ResNet(18) backbone after distillation. As can be seen, our distillation method improves all categories, especially for categories with low accuracy, and reduces the gap between the teacher model and student model after distillation.

\subsection{Results on the Cityscapes Dataset}
\begin{table}[H]
\centering
  \caption{Experimental Results on the Cityscapes Dataset. }
  \label{res:city}
  \scalebox{0.85}{  \begin{tabular}{l|cc|cc}
    \toprule
    Method & val mIoU(\%) & test mIoU(\%) & Params(M) & FLOPs(G) \\
    \midrule
    \midrule
    ENet \cite{paszke2016enet} & - & 58.3 & 0.3580 & 3.612 \\
    ESPNet \cite{Mehta_2018_ECCV} & - & 60.3 & 0.3635 & 4.422 \\
    ERFNet \cite{journals/tits/RomeraABA18} & - & 68.0 & 2.067 & 25.60 \\
    FCN \cite{Long2014Fully} & - & 65.3 & 134.5 & 333.9 \\
    RefineNet \cite{Lin_2017_CVPR} & - & 73.6 & 118.1 & 525.7 \\
    PSPNet \cite{Zhao_2017_CVPR} & - & 78.4 & 70.43 & 574.9 \\
    \midrule
    Teacher: PSPNet(ResNet101) & 78.56  & 76.78  & 70.43 & 574.9  \\
    \midrule
    ResNet18(0.5) &  55.40 & 54.10  & 3.27 & 31.53  \\
    ResNet18(0.5) + SKD \cite{Liu_2019_CVPR} & 61.60  &  60.50 & 3.27 & 31.53\\  
    ResNet18(0.5) + IFVD \cite{wang2020ifvd} & 63.35  &  63.68 & 3.27 & 31.53\\ 
    ResNet18(0.5) + DRD & 62.92  &  62.87 & 3.27 & 31.53\\   
    \midrule
    ResNet18 & 69.10 & 67.60 & 13.07 & 125.8  \\
    ResNet18 + SKD \cite{Liu_2019_CVPR} & 72.70  & 71.40 & 13.07 & 125.8\\
    ResNet18 + IFVD \cite{wang2020ifvd} & 74.54  &  72.74 & 13.07 & 125.8\\
    ResNet18 + DRD & 73.47  &  72.75 & 13.07 & 125.8\\ 
  \bottomrule
\end{tabular}}
\end{table}

\begin{figure}[H]
  \centering
  \includegraphics[width=1.00\linewidth]{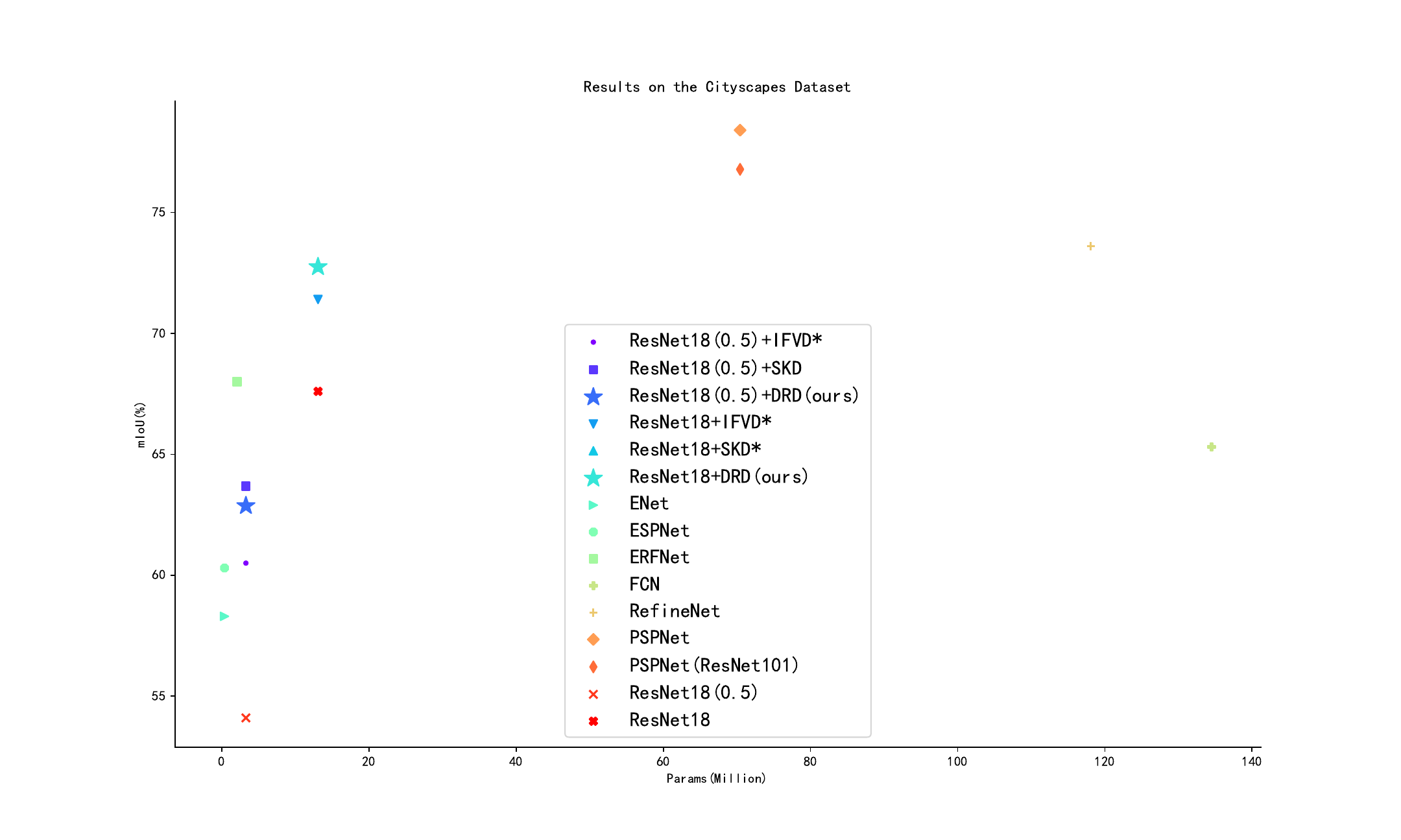}
  \caption{Performance (mIoU (\%)) vs. models sizes of some methods in Table~\ref{res:city}  on Cityscapes Dataset.}
  \label{pic_cityscapes}
\end{figure}

To test the generalization performance of our network, we conduct further experiments on the Cityscapes dataset, a popular benchmark for general scene segmentation. The numerical results are shown in Table~\ref{res:city}, which includes the total segmentation accuracy measured by mIoU, model size, and model complexity.

Our distillation approach improves the segmentation results on both the validation set and test set for two compact networks: ResNet18 (0.5) and ResNet18. Specifically, our DRD improves the student model using the ResNet18 backbone by 4.37\% on the validation set and 4.49\% on the test set after distillation. When we set the backbone as ResNet18 (0.5), a width-halved version of the original ResNet18 that is not pre-trained on ImageNet, our DRD leads to an improvement of 7.52\% on the validation set and 8.77\% on the test set. Compared to previous work, our DRD generates better or comparable results. To clearly demonstrate the performance and model size correlation between the pre and post distillation models, as well as the compact and cumbersome models, we have visually summarized some methods in Table~\ref{res:city} in Figure ~\ref{pic_cityscapes}


\begin{figure}[H]
  \centering
  \includegraphics[width=\linewidth]{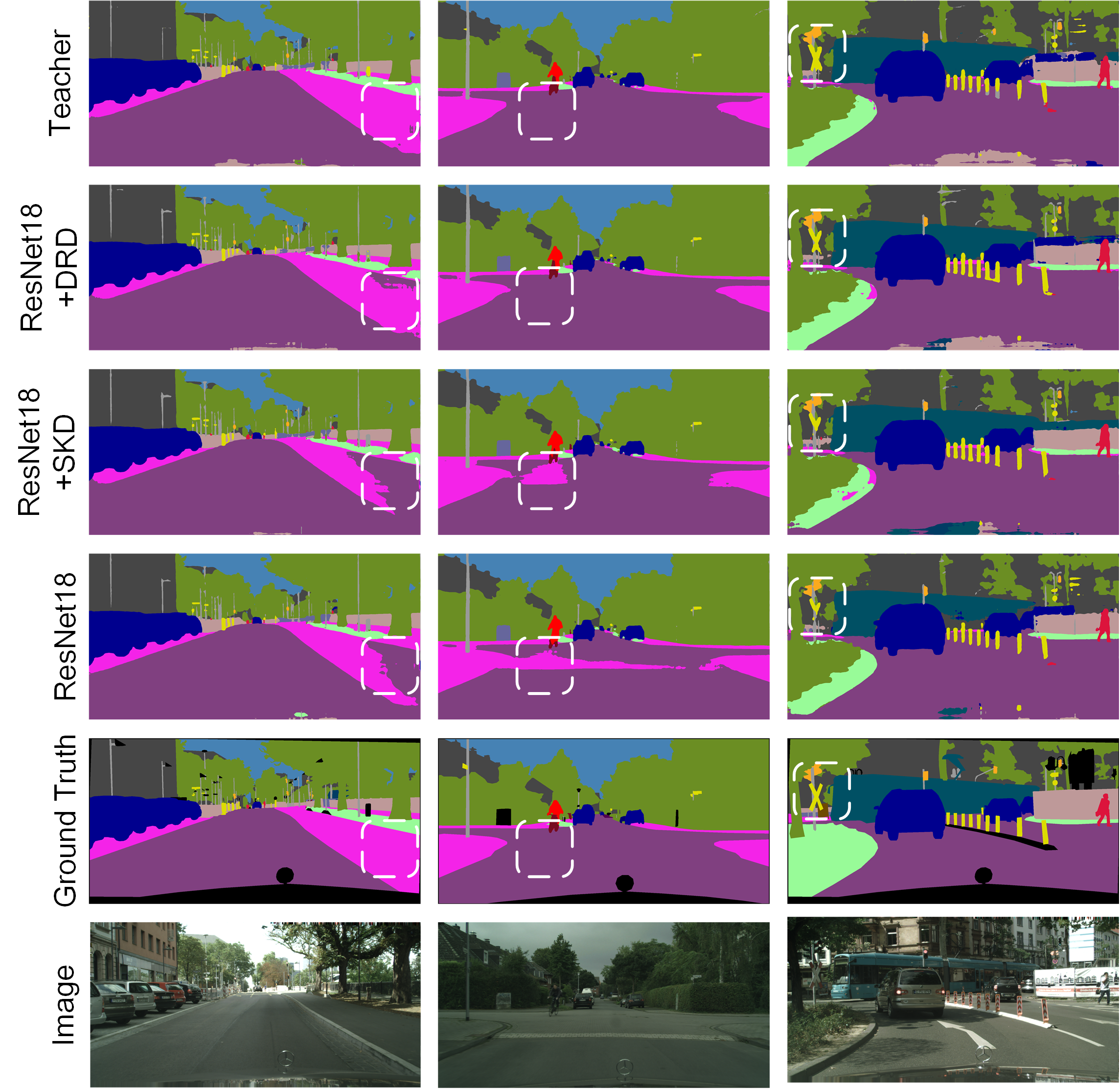}
  \caption{Examples of segmentation results on the Cityscapes dataset.(Best viewed in color.)}
  \label{pic:cityres}
\end{figure}
Figure~\ref{pic:cityres} shows a few examples of qualitative results on the Cityscapes dataset. In the figure, we use dashed rectangular boxes to highlight the areas that require special attention. The segmentation results in these areas can reflect the differences in segmentation results among different methods. If we compare the top row with row 1, row 2, row 3 and row 4, we can conclude that the gap between the student model and teacher model is reduced after distillation, and the student network generates more accurate segmentation results compared to no extra guidance from the teacher network. By comparing row 2 and row 3, we can see that our DRD leads to better segmentation results compared to \cite{Liu_2019_CVPR}.}

\section{Conclusion}
\label{sec:conclusion}

{ In this paper, we have investigated knowledge distillation for training compact semantic segmentation networks with the assistance of cumbersome networks. We have proposed a novel dual relation distillation framework to transfer rich correlation information in feature maps from the teacher network to the student network. These correlations help the student mimic the teacher better in terms of feature distribution, thereby improving the segmentation accuracy of the student model. The comprehensive experimental results have shown that the proposed distillation framework has achieved better or comparable performance compared to previous state-of-the-art methods  on three benchmark datasets, including aerial scenes and general scenes. We believe that our work can benefit others in the field by providing a new approach to training compact semantic segmentation networks with high accuracy. In the future, we aim to explore the use of other types of networks, such as attention-based networks, to improve the performance of our method.}

\section*{Conflict of Interest Statement}
We declare that we have no financial or personal relationships with other people or organizations that could inappropriately influence (bias) this work.

\section*{Acknowledgment}
This work is supported by  NSFC Key Projects of International (Regional) Cooperation and Exchanges under Grant 61860206004, and  NSFC projects under Grant 61976201.

\bibliography{mybibfile}

\end{document}